\documentclass[a4paper,12pt]{article}

\usepackage{amsmath,amssymb,amsfonts} 
\usepackage{graphicx}                 
\usepackage{color}                   
\usepackage{hyperref}     
\definecolor{red}{rgb}{1,0,0}           
\definecolor{green}{rgb}{0,1,0}
\definecolor{blue}{rgb}{0,0,1}
\definecolor{darkblue}{rgb}{0,0,0.5}
\definecolor{lightblue}{rgb}{.5,.5,1}
\definecolor{lightgray}{gray}{.87}          
\definecolor{Dark}{gray}{.20}
\definecolor{pink}{rgb}{.95,0.82,0.92}  
\definecolor{yellow}{rgb}{1,1,0}
\definecolor{lightyellow}{rgb}{1,1,.5}
\definecolor{purple}{rgb}{0.7,0,0.85}
\definecolor{darkgreen}{rgb}{0,0.5,0}
\definecolor{orange}{rgb}{0.8,0.2,0.2}
\def \be {\begin{equation}}
\def \ee {\end{equation}}
\def \bea {\begin{eqnarray}}
\def \eea {\end{eqnarray}}
\def \nn {\nonumber}

\def \rr {\raise.35ex\hbox{\small $\prime$}\kern-.17em{\mbox{\large $\imath$}}}
\def \del {\partial}
\def \dels {\partial\kern-.5em / \kern.5em}
\def \As {{A\kern-.5em / \kern.5em}}
\def \Ds {D\kern-.7em / \kern.5em}

\def \lam {\lambda}

\def \r {\xi}

\def \th {\theta}

\def \k {\lambda}

\newcommand{\detail}[1]{}

\newcommand{\hide}[1]{}

\begin{document}

\pagestyle{plain}


\begin{titlepage}

\begin{center}

\noindent
\textbf{\LARGE
\vskip2cm
On the Near-Horizon Geometry \\
\vskip10pt
of an Evaporating Black Hole
}
\vskip .5in
{\large 
Pei-Ming Ho
\footnote{e-mail address: pmho@phys.ntu.edu.tw},
Yoshinori Matsuo
\footnote{e-mail address: matsuo@phys.ntu.edu.tw}
}
\\
{\vskip 10mm \sl

Department of Physics and Center for Theoretical Physics, \\
National Taiwan University, Taipei 106, Taiwan,
R.O.C. 
}\\

\vspace{60pt}
\begin{abstract}

The near-horizon geometry of evaporation black holes
is determined according to the semi-classical Einstein equation.
We consider spherically symmetric configurations
in which the collapsing star has already collapsed below
the Schwarzschild radius.
The back-reaction of the vacuum energy-momentum,
including Hawking radiation,
is taken into account.
The vacuum energy-momentum plays a crucial role 
in a small neighborhood of the apparent horizon,
as it appears at the leading order 
in the semi-classical Einstein equation.
Our study is focused on the time-dependent geometry in this region.

\end{abstract}
\end{center}

\end{titlepage}

\setcounter{page}{1}
\setcounter{footnote}{0}
\setcounter{section}{0}


\section{Introduction}

Black holes are found in the universe over a wide range of masses.
There are black holes only a few times heavier than the sun
as well as supermassive black holes that are several billion solar masses.
It is possible that they are formed through different dynamical processes,
and that they behave differently as astrophysical black holes.
Hence it is of interest to explore all theoretically viable
black-hole-like configurations.
In this work,
we present new scenarios for black-hole formation and evaporation
which were largely overlooked in the literature.

Conventionally,
it is assumed that,
when a star is sufficiently compact,
the gravitational force dominates over all other forces,
so the star continues to collapse until it is of Planckian size.
A theoretical basis for this assumption is the Buchdahl theorem
\cite{Buchdahl:1959zz}.
It states that 
any static configuration suffers a divergence in pressure 
if its radius is smaller than $9/4$ of its mass
($9/8$ of its Schwarzschild radius). 

However, 
the Buchdahl theorem assumes the weak energy condition,
which is typically violated by quantum effects.
With the vacuum energy-momentum tensor taken into account,
there can be regular static solutions 
for which the Buchdahl theorem is inapplicable
\cite{Ho:2017joh,Ho:2017vgi}.
Due to the back reaction of the vacuum energy-momentum tensor,
the geometry around the Schwarzschild radius is wormhole-like
\cite{Fabbri,Ho:2017joh,Ho:2017vgi,Berthiere:2017tms},
i.e.,
it has a local minimum in the areal radius $r$,
but the static solution is horizonless. 

The possibility of black-hole-like configurations 
with wormhole-like structures was proposed over 10 years ago
\cite{wormhole-as-BH,Fabbri}.
Recently,
in Ref.\cite{Ho:2017joh,Ho:2017vgi},
this idea was realized in a static solution to 
the semi-classical Einstein equation,
which adopted the vacuum energy-momentum tensor
derived from a 2D quantum field theory \cite{Davies:1976ei}.
A similar horizonless wormhole-like solution was 
later also found for the vacuum energy-momentum tensor of
a 4D quantum field theory in Ref.\cite{Berthiere:2017tms}.
While the wormhole-like structure might be 
a generic feature for a certain class of quantum field theories,
it is of interest to understand how 
these wormhole-like structures are modified 
when the time-dependence is turned on.

In this paper,
we study the dynamical solutions of
the semi-classical Einstein equation.
We consider the vacuum energy-momentum tensor
proposed in Ref.\cite{Davies:1976ei},
so the dynamical solutions can be compared with 
the static solutions found in Refs.\cite{Ho:2017joh,Ho:2017vgi}.
In fact, the black holes defined by the same equations
have already been considered in Ref.\cite{Davies:1976ei},
where the Schwarzschild metric was assumed to be
the 0-th order approximation.
But since the black hole eventually completely evaporates,
the Schwarzschild radius must be allowed to change with time
in order for this approximation to have any chance to be valid.
However,
even the time-dependent Schwarzschild metric
can not be a good approximation
for the geometry close to the Schwarzschild radius.
It was shown
\cite{Kawai:2013mda,Kawai:2014afa,Ho:2015fja,Kawai:2015uya,Ho:2015vga}
that the time-dependent Schwarzschild metric does not
allow any infalling causal trajectory to cross 
the Schwarzschild radius,
assuming that the Schwarzschild radius monotonically 
shrinks to zero within finite time,
regardless of how slowly the Schwarzschild radius shrinks.
As previous works on the near-horizon geometry is unreliable,
we shall thus focus on the geometry around the Schwarzschild radius,
and study the time-dependence of the wormhole-like geometry.
We will comment on the geometry inside the Schwarzschild radius
at the end of this paper.

We will study the near-horizon geometry via 
two different perturbation theories.
We will first
consider the approximation in which 
the exterior geometry of the black hole 
changes very slowly with time
--- as it is expected since
the Hawking radiation is extremely weak
--- so that,
at any instant of time,
the static solution
(with time-dependent parameters 
that change extremely slowly over time)
can be viewed as the 0-th order approximation.
The back reaction of the vacuum energy-momentum tensor to 
the space-time geometry is included in the 0-th order solution.
The second perturbative formulation is the $\hbar$-expansion
(or equivalently the expansion of Newton constant).
We will show that the vacuum energy-momentum tensor
(including Hawking radiation) appears at the leading order
in the semi-classical Einstein equation
when we focus on the near-horizon geometry.

In our analysis,
the apparent horizon can appear
when the Hawking radiation is turned on,
while the wormhole-like structure persists. 
The apparent horizon,
unlike the horizon for static solutions,
is timelike due to the negative vacuum energy,
and hence the information can get away from the trapped region. 
The rate at which the minimal areal radius decreases
is proportional to the strength of the Hawking radiation,
satisfying a formula resembling
its conventional-model counterpart
between the Schwarzschild radius and the Hawking radiation. 
The geometry near the apparent horizon resembles 
the conventional model for black-hole evaporation,
although the size of the space inside the apparent horizon can be different. 

The plan of this paper is the following.
We define in Sec.\ref{EE}
the semi-classical Einstein equation 
that is the basis of our analysis,
and review in Sec.\ref{static-solution}
the results of Refs.\cite{Ho:2017joh,Ho:2017vgi}
on the static solutions with the wormhole-like structure.
In Sec.\ref{derivative-expansion},
we use the derivative expansion to define a perturbation theory 
to deal with the extremely weak time dependence 
due to the Hawking radiation,
in particular,
to determine the time-dependence of the wormhole-like structure.
In Sec.\ref{near-Schwarzschild},
we extend our result
by adopting the
perturbative expansion of $\hbar$
(or equivalently the Newton constant $\kappa$)
in which we zoom into a small neighborhood 
of the neck of the wormhole-like structure.
Finally,
we comment in Sec.\ref{Conclusion}
on the implications of the results of our analysis.

\section{Semi-Classical Einstein Equation}
\label{EE}

In this section,
we define the semi-classical Einstein equation
and review the model for the vacuum energy-momentum tensor 
proposed in Ref.\cite{Davies:1976ei}.

Assuming that the spacetime has spherical symmetry,
the metric can be written in the form
\be
ds^2 = - C(u, v) du dv + r^2(u, v) d\Omega^2,
\label{metric}
\ee
where $(u, v)$ are light-like coordinates and 
$d\Omega^2 = d\theta^2 + \sin^2\th d\phi^2$
is the line element of a unit 2-sphere.
We will only consider asymptotically flat spacetimes in this paper.
As a convention,
we choose $(u, v)$ such that $C(u, v) \rightarrow 1$ as $r \rightarrow \infty$.
The Minkowski time $t$ at spatial infinity is related to $(u, v)$ via
$u = t - r$ and $v = t + r$.

The Einstein tensor for the metric (\ref{metric}) is given by
\begin{align}
G_{uu} &= \frac{2\del_u C\del_u r}{Cr}- \frac{2\del_u^2 r}{r}, \\
G_{vv} &= \frac{2\del_v C\del_v r}{Cr}- \frac{2\del_v^2 r}{r}, \\
G_{uv} &= \frac{C}{2 r^2} + \frac{2\del_u r\del_v r}{r^2} + \frac{2\del_u\del_v r}{r}, 
\\
G_{\theta\theta} &= 
\frac{2 r^2 \left(\partial_u C \partial_v C - C \partial_u \partial_v C\right)}{C^3} 
- \frac{4r \partial_u \partial_v r}{C},
\end{align}
and $G_{\phi\phi}$ equals $G_{\th\th}$ up to an overall factor of $\sin^2\th$.

The evaluation of the vacuum expectation value of the quantum energy-momentum operator 
in an arbitrary background of spacetime is a very complicated task.
We follow Ref.\cite{Davies:1976ei} to adopt the quantum theory of 2D massless scalar fields as
a dimensionally reduced model with spherical symmetry
for the vacuum energy-momentum tensor.
We focus on the spherically symmetric configurations, 
and assume that the energy-momentum tensor satisfies 
the 2D conservation law after integrating over the angular directions. 
Furthermore, we assume that the energy-momentum tensor 
has the Weyl anomaly for 2D scalar fields. 
This model is widely used in the literature
\cite{Barcelo:2007yk,Christensen:1977jc,Parentani:1994ij,Brout:1995rd,Ayal:1997ab,Fabbri}.
The energy-momentum tensor is fixed by the 2D conservation law and 
anomaly condition up to integration constants. 
Then, the 4D vacuum energy-momentum tensor is given by
\begin{align}
\langle T_{uu} \rangle &=
- \frac{N}{12\pi r^2} C^{1/2} \del_u^2 C^{-1/2} + 
\frac{
f_u(u)
}{r^2},
\label{Tuu-vac}
\\
\langle T_{vv} \rangle &=
- \frac{N}{12\pi r^2} C^{1/2} \del_v^2 C^{-1/2} + \frac{f_v(v)}{r^2},
\label{Tvv-vac}
\\
\langle T_{uv} \rangle &=
- \frac{N}{24\pi r^2 C^2}
\left[ C \del_u\del_v C - \del_u C \del_v C \right],
\label{Tuv-vac}
\\
\langle T_{\th\th} \rangle &= 0 \ . 
\label{Tthth-vac}
\end{align}

It should be noted that $\langle T_{\th\th} \rangle$ is zero
because in this model 
the energy-momentum tensor satisfies both the 2D and 4D conservation laws, 
but would be non-zero for more general energy-momentum 
which satisfies only the 4D conservation law.
The functions $f_u(u)$ and $f_v(v)$ comes from
the integration constants for the conservation law. 
They correspond to the outgoing and ingoing energy fluxes
at $r \rightarrow \infty$ 
if $(u,v)$-coordinates are chosen such that 
$C\to$ constant in $r\to\infty$. 
It should also be noted that the energy-momentum tensor 
is independent of the physical state of matter fields 
except for $f_u(u)$ and $f_v(v)$, 
since the Weyl anomaly is independent of the physical state of matters. 
Therefore, all information on matters are contained in $f_u(u)$ and $f_v(v)$, 
and other terms have no information. 
Here $N$ is the number of 2D scalar fields 
contributing to the vacuum energy-momentum tensor.
For convenience, 
we define the parameter
\be
\alpha \equiv \frac{\kappa N}{24\pi},
\ee
where $\kappa$ is the Newton constant times $8\pi$.
The parameter $\alpha$ characterizes the magnitude of 
the contribution of the vacuum energy-momentum tensor 
to the semi-classical Einstein equation.

We assume in this paper that 
the semi-classical Einstein equation 
\be
G_{\mu\nu} = \kappa \langle T_{\mu\nu} \rangle
\ee
properly account for the back reaction of the vacuum energy-momentum tensor 
to the classical geometry of spacetime
when the curvature is much smaller than the Planck scale.

For this model, 
it is shown that static solutions have no horizon 
in the absence of outgoing or ingoing energy fluxes
($f_u(u) = f_v(v) = 0$) \cite{Ho:2017joh}. 
The static configuration of a star composed of the perfect fluid
was also studied as a solution 
to the semi-classical Einstein equation, 
by introducing the classical energy-momentum tensor 
of the fluid in addition to the vacuum energy-momentum tensor above \cite{Ho:2017vgi}.
It was found that the vacuum energy-momentum tensor of this toy model
automatically regularizes the geometry such that 
$C(u, v)$ never goes to $0$ and the pressure $T_{\th\th}$ never diverges.
The Buchdahl theorem is circumvented as 
the quantum energy in vacuum can violate the weak energy condition.

\section{Static Solution}
\label{static-solution}

Let us now briefly review the static solution to
the semi-classical Einstein equation 
\cite{Ho:2017joh,Ho:2017vgi}, 
assuming that the surface of the star 
is under the neck of the wormhole-like structure
We shall focus on the vacuum geometry around the neck 
in the static solution in this section.

The static solution to the semi-classical Einstein equation is 
found in this form 
\be
ds^2 = - C(r_*) (dt^2 - dr_*^2) + r^2(r_*) d\Omega^2 \ . 
\ee
For an asymptotically Minskowskian spherically symmetric static star,
we expect that
the geometry is well approximated by the Schwarzschild metric,
that is,
\begin{align}
C(r_*) &\simeq 1 - \frac{a_0}{r(r_*)},
\\
r_* &\simeq r + a_0 \log\left(\frac{r}{a_0} - 1\right)
\end{align}
for $r_* \rightarrow \infty$,
where $a_0$ is the Schwarzschild radius.

We will refer to $r$ as the areal radius 
since the 2-sphere defined by constant $r$
(for any fixed $t$)
has the area $4\pi r^2$.
In a curved space,
the proper distance from the origin to 
a point on the 2-sphere with areal radius $r$
is in general not the same as $r$.
We shall refer to this proper distance as 
the radial proper distance.

Defining a function $F(r)$ by
\be
\frac{dr}{dr_*} = F(r),
\ee
one can in principle derive the function $r(r_*)$ for given $F(r)$.
A spherically symmetric metric is therefore specified by
two functions $C(r_*)$ and $F(r_*)$.
Note that in general $F(r_*)$ is not positive-definite.
It means that
the areal radius $r$ is not monotonically increasing 
as we move away from the origin in the radial direction,
unlike the coordinate $r_*$.
The function $r_*(r)$ is thus possibly multiple-valued
and should be defined separately for different 
branches of $r$
(divided by local extrema of $r$)
on which $r_*(r)$ is monotonic.

The event horizon for a static configuration is located where $C(r_*) = 0$.
It was proven analytically \cite{Ho:2017joh} that
the model of vacuum energy described above does not admit an event horizon.
There can be no apparent horizon either because 
apparent horizons coincide with the event horizon for static configurations.

\begin{figure}
\vskip1cm
\begin{center}
\includegraphics[scale=0.5,bb=0 0 375 280]{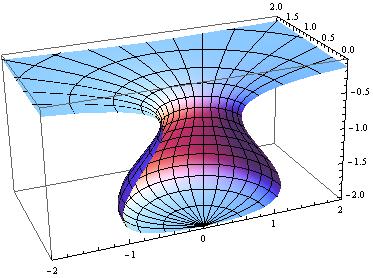}
\caption{\small 
The schematic sectional view of the static solution to 
the semi-classical Einstein equation.
The neck is the local minimum of the areal radius $r$.
The surface of the star is roughly where $r$ has a local maximum.
}
\label{fig:near-neck-vacuum}
\end{center}
\end{figure}

What replaces the horizon at the Schwarzschild radius $a_0 = 2M$
is the ``neck,''
i.e.\ a minimum in the areal radius $r$
at the quantum Schwarzschild radius $a$
whose deviation from the classical Schwarzschild radius $a_0$ 
is no larger than $\mathcal O(\alpha/a)$. 
Under the neck,
the areal radius $r$ increases as we move towards the origin,
until we enter the interior space of the star.
(Inside the star, 
the areal radius $r$ goes to zero at the origin.)
A schematic sectional view of the geometry of such a static configuration
is given in Fig.\ref{fig:near-neck-vacuum}. 

While numerical solutions of the two functions $C(r)$ and $F(r)$ 
are calculated,
analytic expressions are obtained by using 
the expansion around the neck
or in asymptotic regions \cite{Ho:2017joh,Ho:2017vgi}. 
For the space-time region above the neck with a separation
(proper distance) much larger than the Planck length
(which implies that $r - a \gg \frac{\alpha}{a}$), 
the Schwarzschild solution is a good approximation. 
As one further approaches to the Schwarzschild radius, 
the Schwarzschild solution no longer gives a good approximation. 
In the very small region in which
\be
r - a \ll \frac{\alpha}{a}
\ee
(all points in this region are within a proper distance of 
a Planck length or less away from the neck),
we have the following analytical expressions as a good approximation 
of a solution of the semi-classical Einstein equation 
\begin{align}
C(r) \simeq c_0 \, e^{2\sqrt{k(r-a)}},
\qquad
F(r) \simeq \sqrt{\frac{4 c_0(r-a)}{\alpha k}}
\simeq \sqrt{\frac{2 c_0(r-a)}{a}},
\end{align}
where
\be
k = \frac{2(a^2 - \alpha)}{\alpha a} \simeq \frac{2a}{\alpha}.
\ee
The metric here resembles that of a traversable wormhole
and it is approximately given by
\begin{align}
ds^2 &\simeq - c_0 e^{2\sqrt{k(r-a)}} dt^2 + \frac{a}{2(r-a)} dr^2 + r^2 d\Omega^2
\nn \\
&\simeq - \left[c_0 + 2\sqrt{\frac{c_0}{\alpha}}x\right] (dt^2 - dx^2)
+ \left[a^2 + c_0 x^2\right] d\Omega^2,
\label{metric-neck}
\end{align}
where the radial coordinate $x$ is related to $r$ via%
\footnote{
In comparison with the notation used in Ref.\cite{Ho:2017joh},
$x = r_* - a_*$
for some constant $a_*$.
}
\be
r \simeq a + \frac{c_0}{2 a} x^2 + \cdots.
\ee
Since this metric is connected to the Schwarzschild metric
with $C = 1 - a_0/r$
around the region where $r - a \sim \mathcal O(\alpha/a)$ above the neck,
the parameter $c_0$ must be of order 
\begin{equation}
c_0 = {\cal O}(\alpha/a^2) \ .
\label{c0-estimate-0}
\end{equation}

The coordinate $x$ is defined such that it is a monotonic function 
of the radial direction.
For the space above (below) the neck,
$x > 0$ ($x < 0$),
the areal radius $r$ increases as we move away from the neck 
regardless of whether we are moving inward or outward.
The metric above \eqref{metric-neck} applies to $r - a \ll \alpha/a$.

If the surface of the star is under the neck but 
is not in the region $r - a \ll \alpha/a$, 
the geometry is approximated by another asymptotic expression 
for the region under the neck with $r - a \gg \alpha/a$
(but before reaching the surface of the star) 
\begin{align}
C(r) &= e^{2\rho(r)},
\quad 
\mbox{where}
\quad
\frac{d}{dr}\rho(r) \simeq - \frac{2r}{\alpha}
-  2 f_0 r^2 e^{-2r^2/\alpha},
\\
F(r) &\simeq - \frac{1}{f_0^{1/2} r},
\end{align}
where the parameter $f_0 > 0$ is bounded from above by the requirement that
the second term in $\frac{d\rho}{dr}$ is much smaller than the first term,
i.e. $f_0 e^{-2a^2/\alpha} \ll \frac{1}{\alpha a}$.

The corresponding metric is approximately
\be
ds^2 \simeq - r e^{-2r^2/\alpha} \left[ \tilde{c}_0 dt^2 - f_0 r^2 dr^2 \right] + r^2 d\Omega^2,
\ee
where $\tilde{c}_0$ is of order
\be
\tilde{c}_0 \sim {\cal O}\left( e^{2a^2/\alpha} \frac{\alpha}{a^3} \right).
\ee
The values of the parameters $f_0$ and $\tilde{c}_0$ 
can be fixed by connecting the metric in this region
with that in other regions.
The proper length of this region in the radial direction is 
typically of the order of Planck scale. 
However, if the surface of the star is under the neck 
but in the region $r - a \gg \alpha/a$, 
the energy density of the star must be very large,
even beyond the Planck scale. 
This implies that the surface of the compact star
is always very close to the neck. 

In the following,
we will study the modification of the metric \eqref{metric-neck}
around the neck due to Hawking radiation.

\section{Derivative Expansion}
\label{derivative-expansion}

Given the static configurations of the wormhole-like black holes,
we can consider a slow evaporation 
of the star which is induced by an extremely weak Hawking radiation.
This slowly changing geometry
would be well approximated by a static solution at any given moment.
We can solve the semi-classical Einstein equation 
for this slow time-dependent ansatz 
by treating the Hawking radiation perturbatively. 

In this section,
we consider this perturbation theory.
Since the time-dependence of the geometry is a result 
of the back reaction of the Hawking radiation, 
which is a first order perturbation, 
each time derivative acting on a geometric quantity 
increases its order of perturbation by one. 
The perturbative expansion in this theory 
thus involves a derivative expansion. 

Here we use the $(u,x)$-coordinates, 
where $x$ is defined by 
\begin{equation}
 x \equiv \frac{1}{2}\left(v-u\right) \ , 
 \label{x-def}
\end{equation}
up to a constant 
and is an analogue of 
the radial coordinate $x$ in the static solution,
which depends only on $x$.
With the assumption of spherical symmetry,
if there is no outgoing energy flux,
the static solution would still apply to
the space-time outside the star
during the gravitational collapse.
But due to the Hawking radiation,
the solution to the semi-classical Einstein equation 
has a mild $u$-dependence.

To define the perturbation theory for the mild $u$-dependence,
we introduce the expansion parameter $\epsilon$.
The Hawking radiation $\beta$, 
which appears in $T_{uu}$ as 
\begin{align}
T_{uu} &=
- \frac{1}{12\pi r^2} C^{1/2} \del_u^2 C^{-1/2} + \frac{\beta(u)}
{24\pi r^2} \ ,
\label{Tuu-beta}
\end{align}
is by definition of the first order
\begin{equation}
 \beta(u) = \mathcal O(\epsilon) \ . 
\end{equation}
It is the origin of the $u$-dependence of the solution.
Correspondingly,
the time derivatives are of the first order:
\begin{equation}
 \partial_u = \mathcal O(\epsilon) \ . 
\end{equation}

Then we expand the semi-classical Einstein equation
with respect to $\epsilon$, 
in which $C(u,x)$ and $r(u,x)$ are expansions of $\epsilon$ as 
\begin{align}
 C(u,x) &= C_0(u,x) + \epsilon C_1(u,x) + \epsilon^2 C_2(u,x) + \cdots \ , 
 \label{C-expand-eps}
\\
 r(u,x) &= r_0(u,x) + \epsilon r_1(u,x) + \epsilon^2 r_2(u,x) + \cdots \ .  
 \label{r-expand-eps}
\end{align}
The leading order terms of the semi-classical Einstein equation 
give the static solution. 
The Hawking radiation $\beta$ and $u$-derivatives are of $\mathcal O(\epsilon)$
and do not appear in the leading order terms. 

Define $\rho_0$ by
\begin{equation}
 C_0(u,x) = e^{2 \rho_0(u,x)}\ .
\end{equation}
The semi-classical Einstein equation at the 0-th order is equivalent to
the following two independent differential equations:
\begin{align}
 0 &= 
 r_0 \left[2 \left(\partial_x r_0\right) \left(\partial_x \rho_0\right) 
 - \partial_x^2 r_0\right] 
 + \alpha \left[\left(\partial_x \rho_0\right)^2 
 - \partial_x^2 \rho_0\right] \ , 
\label{eqvv00}
\\
 0 &= 
 e^{2 \rho_0} + \left(\partial_x r_0\right)^2 
 + r_0 \partial_x^2 r_0 + \alpha \partial_x^2 \rho_0 \ . 
\label{equv00}
\end{align}
These differential equations have a trivial solution; 
\begin{align}
 \rho_0 (u,x) &= 
 \frac{1}{2}\log \alpha - \log\left(x-x_0(u)\right) \ , 
\label{trivial-rho}
\\
 r_0(u,x) &= a(u) \ . 
\label{trivial_r}
\end{align}
where $x_0(u)$ and $a(u)$ are integration constants for $x$-integrals. 
This solution is clearly not describing a black-hole geometry
since $r_0$ is completely independent of the radial coordinate $x$. 

Now we look for solutions other than the trivial solution.
Since it is difficult to solve the semi-classical Einstein equations
\eqref{eqvv00} and \eqref{equv00} exactly, 
we consider the $x$-expansion around the Schwarzschild radius. 
Notice that the differential equations \eqref{eqvv00} and \eqref{equv00} 
contain second order derivatives of $x$ only through the combination of 
$(r_0 \partial_x^2 r_0 + \alpha \partial_x^2 \rho_0)$, 
hence the solution in general has only 3,
rather than 4, integration constants. 
(They are integration constants for $x$-integrals 
so they have $u$-dependence.)

We expand $\rho_0$ and $r_0$ around $x=0$ in powers of $x$ as 
\begin{align}
 \rho_0(u,x) &= \frac{1}{2} \log c_0(u) + \rho_{01}(u) x + \rho_{02}(u) x^2 + \cdots \ , 
\\
 r_0 (u,x) &= a(u) + r_{01}(u) x + r_{02}(u) x^2 + \cdots \ .
\end{align}
The functions $c_0(u)$ and $a(u)$ in these expansions 
are expected to be two of the integration constants mentioned above.
In principle,
they can be fixed by
the asymptotic behavior of the solution.

The third integration constant can be identified with a shift in $x$,
which is a symmetry of the differential equations \eqref{eqvv00}, \eqref{equv00}.
Since it is already known \cite{Ho:2017joh} that there is a local minimum of $r$ 
at the quantum Schwarzschild radius, 
we choose to fix the third integration constant such that $x=0$
at the local minimum of $r$, 
namely, 
\begin{equation}
r_{01}(u) = 0 \ .
\end{equation} 

The solution to the differential equations \eqref{eqvv00} and \eqref{equv00}
can now be solved as expansions in powers of $x$ as 
\begin{align}
 \rho_0(u,x) &= \frac{1}{2} \log c_0(u) + \alpha^{-1/2} c_0^{1/2}(u) x 
 - \frac{c_0(u)}{2\left(a^2(u) - \alpha\right)} x^2 + \mathcal O(x^3)  \ , 
\label{rho0-sol}
\\
 r_0(u,x) 
 &= 
 a(u) + \frac{a(u) c_0(u)}{2\left(a^2(u) - \alpha\right)} x^2 + \mathcal O(x^3) \ . 
\label{r0-sol}
\end{align}
This is essentially a promotion of the static solution around the neck
to allow $u$-dependence in the coefficients.
The parametric functions $a(u)$ and $c_0(u)$ are expected 
to change very slowly with time.
Over a relatively short period of time
and within a relatively small region of space,
the geometry is well approximated by the static solution.

Let us now consider the semi-classical Einstein equation
at the next-to-leading order
in the perturbation theory of the $\epsilon$-expansion.
At this order,
the semi-classical Einstein equation leads to
3 independent differential equations. 
They can be organized such that
two of them are the differential equations for 
the next-to-leading order terms $C_1$ and $r_1$
in the expansions \eqref{C-expand-eps}, \eqref{r-expand-eps},
with the third equation involveing only 
$\beta$ and the 0-th order terms $a(u)$ and $c_0(u)$,
but not $C_1$ nor $r_1$.
The 3rd equation serves as a constraint on the parameters | 
the Hawking radiation parameter $\beta$ 
and the integration constants $c_0(u)$ and $a(u)$. 
This is the equation we are most interested in.

This constraint equation can be derived by
taking the difference between 
the $\mathcal O(\epsilon)$ terms in the $(u,u)$-component 
and those in the $(v,v)$-component 
of the semi-classical Einstein equation.
It is
\begin{equation}
 0 = 
 r_0 \left[\left(\partial_u r_0\right) \left(\partial_x \rho_0\right) 
 + \left(\partial_x r_0\right) \left(\partial_u \rho_0\right) - \partial_x \partial_u r_0\right] 
 + \alpha \left[ \frac{1}{2} \beta + 
  \left(\partial_u \rho_0\right) \left(\partial_x \rho_0\right) - \partial_x \partial_u \rho_0\right] \ . 
\end{equation}
This constraint relates the Hawking radiation parameter $\beta$
to the first time derivative of the quantum Schwarzschild radius $\dot a$. 
Substituting the solution 
\eqref{rho0-sol} and \eqref{r0-sol} at the leading order 
in this constraint equation,
we obtain 
\begin{equation}
 \dot a(u) = - \frac{\alpha^{3/2}\beta(u)}{2 c_0^{1/2}(u) a(u)} \ . 
 \label{dota-1}
\end{equation}

In order for the 0-th order solution 
to be continued to the asymptotic time-dependent Schwarzschild solution
(more precisely, the outgoing Vaidya solution)
at large distances,
we find 
\footnote{
The derivation of eq.\eqref{c0-estimate} is essentially
the same as that of eq.\eqref{c0-estimate-0}.
}
\begin{equation}
 c_0(u) \sim \mathcal O\left(\frac{\alpha}{a^2(u)}\right) \ ,
 \label{c0-estimate}
\end{equation}
and thus
\begin{equation}
 \dot a(u) \sim - \mathcal O\left(\alpha\beta(u)\right) \ . 
 \label{a-beta-1}
\end{equation}

Since the Hawking radiation determines 
the loss of energy at large distances,
and the space-time geometry far from the collapsing sphere
is approximated by the Schwarzschild metric,
the classical Schwarzschild radius $a_0$ decreases at the rate
\begin{equation}
\dot a_0(u) \simeq - \alpha \beta \ .
\label{dota0-1}
\end{equation}
While $a(u)$ and $a_0(u)$ differ only by $\mathcal O(\alpha^{1/2})$, i.e. Planck-length scale, 
our result eq.\eqref{dota-1} should agree with \eqref{dota0-1} 
at the leading order of $\alpha$-expansion. 
Therefore, we would have
\begin{equation}
 c_0(u) \simeq \frac{\alpha}{4 a^2(u)} \ ,
 \label{c0-estimate-1}
\end{equation}
as an improvement of the estimate \eqref{c0-estimate}.

\section{The Geometry Around the Neck}
\label{near-Schwarzschild}

In the previous section,
we considered the derivative expansion of
the semi-classical Einstein equations.
While this perturbation theory is applicable
to the whole space outside the star,
we have assumed that the Hawking radiation is smaller than 
the other quantum effects in the energy-momentum tensor. 
This assumption in fact is not very realistic since 
the Hawking radiation is sometimes comparable to 
the vacuum energy of the static solution
near the Schwarzschild radius. 
In this section,
we shall develop another perturbation theory that is 
only applicable to the small neighborhood
of the neck of the wormhole-like structure.
But instead of the $\epsilon$-expansion 
(i.e.\ the Hawking radiation is much smaller than the static vacuum energy) 
in the previous section,
we consider the expansion in the parameter $\alpha$
(or equivalently $\kappa$). 
As we are still focusing on the region outside the collapsing star,
the only contribution to the energy-momentum tensor is 
the vacuum expectation value of the quantum energy-momentum operator.
Hence the expansion in $\alpha$ is also equivalent to 
the expansion in the parameter $\hbar$ for the quantum effect.

The region we are interested in 
is the small neighborhood within Planck-length distance 
from the neck at the quantum Schwarzchild radius $a$.
That is,
we consider the semi-classical Einstein equation 
near the quantum Schwarzschild radius for%
\footnote{
To be more precise, \eqref{r=a+alpha} should be expressed as 
$r(u,x) = a(u) + \mathcal O\left(\frac{\alpha}{a}\right)$. 
But here, we focus on the power of $\alpha$ and 
the power of $a$ can easily be figure out by dimensional analysis.
}
\begin{equation}
 r(u,x) = a(u) + \mathcal O(\alpha) \ , 
 \label{r=a+alpha}
\end{equation}
where the quantum Schwarzschild radius $a(u)$ depends on $u$
due to the Hawking radiation,
and the spatial coordinate $x$ is defined in eq.\eqref{x-def}.

Since the metric is approximated by the Schwarzschild solution 
\begin{align}
 C(u,x) \simeq 1 - \frac{a(u)}{r(u,x)}
\end{align}
for $r(u,x) - a(u) \gg \mathcal O(\alpha)$,
$C$ is expected to be of $\mathcal O(\alpha)$ for 
$r = a(u) + \mathcal O(\alpha)$. 
Hence, 
in the neighborhood of interest \eqref{r=a+alpha},
we expand $C(u, x)$ and $r(u, x)$ as
\begin{align}
 C(u,x) &= \alpha C_0(u,x) + \alpha^2 C_1(u,x) + \mathcal O(\alpha^3) \ , 
 \label{Cux}
\\
 r(u,x) &= a(u) + \alpha \r_0 (u,x) + \alpha^2 \r_1(u,x) + \mathcal O(\alpha^3) \ ,
 \label{rux}
\end{align}
where $C_0, C_1, C_2, \cdots$ and 
$a, \r_0, \r_1, \cdots$ are the coefficients of the $\alpha$-expansion, 
which are of $\mathcal O(\alpha^0)$.

The Hawking radiation can be expanded as 
\begin{equation}
 \beta(u) = \beta_0(u) + \alpha \beta_1(u) + \mathcal O(\alpha^2) \ ,
\end{equation}
where $\beta_0(u), \beta_1(u), \cdots$ are of $\mathcal O(1)$.
In the semi-classical Einstein equation,
the Hawking radiation contributes a term
of the same order 
as the rest of the vacuum energy-momentum tensor.

If the Hawking radiation is absent,
there would be a static solution depending on $x$ only
(independent of $u$).
The solution becomes $u$-dependent when
the Hawking radiation is turned on.
Therefore,
the $u$-derivatives are expected to be of $\mathcal O(\alpha)$,
namely, 
\begin{equation}
 \partial_u \sim \mathcal O(\alpha) \ .
 \label{delu=Oalpha}
\end{equation}

Despite the derivative expansion suggested by eq.\eqref{delu=Oalpha},
the analysis in this section is different from the previous section
because,
as a result of zooming into the region \eqref{r=a+alpha} at the same time,
the Hawking radiation $\beta$ enters the lowest order terms
in the Einstein equation as we will see below.
This explains why the geometry is modified 
in the region where $r - a \sim \mathcal O(\alpha)$
due to a modification of the energy-momentum tensor of 
$\mathcal O(\alpha)$.

\subsection{Leading Order}

It is straightforward to show that
the semi-classical Einstein equation at the leading order
in the $\alpha$-expansion
is equivalent to the following three equations:
\begin{align}
 0 &= 
 3 \left(\partial_x C_0\right)^2 
 + 4 a C_0 \left(\partial_x C_0\right) \left(\partial_x \r_0\right)  
 - 2 C_0\partial_x^2 C_0  
 - 4 a C_0^2\partial_x^2 \r_0  
\notag\\&\quad
 - 8 \beta_0 C_0^2 - 8 \alpha^{-1} a \dot a C_0 \partial_x C_0 \ , 
 \label{equu}
\\
 0 &= 
 3 \left(\partial_x C_0\right)^2 
 + 4 a C_0 \left(\partial_x C_0\right) \left(\partial_x \r_0\right)  
 - 2 C_0 \partial_x^2 C_0
 - 4 a C_0^2\partial_x^2 \r_0 \ , 
 \label{eqvv}
\\
 0 &= 
 2 C_0^3 + \left(\partial_x C_0\right)^2 - C_0 \partial_x^2 C_0 
 - 2 a C_0^2 \partial_x^2 \r_0 \ , 
 \label{equv}
\end{align}
where $\dot a = \partial_u a \sim \mathcal O(\alpha)$. 
These are clearly the equations for a perturbation theory
different from the previous section,
as we see the Hawking radiation $\beta_0$ appearing at the leading order.
They also demonstrate the fact that the Hawking radiation, 
albeit extremely weak,
can have potentially important back reaction to the spacetime geometry
when we zoom into a sufficiently small neighborhood of the horizon.

By subtracting 2 times eq.\eqref{equv} from eq.\eqref{eqvv}, 
we find
\begin{equation}
 \partial_x \r_0 =
 \frac{4 C_0^3 - \left(\partial_x C_0\right)^2}{4 a C_0 \partial_x C_0} \ . 
 \label{delxr0}
\end{equation}
Substituting this back into \eqref{eqvv} (or \eqref{equv}), 
we obtain 
\begin{equation}
 0 = \left[4 C_0^3 - \left(\partial_x C_0\right)^2\right]
 \left[C_0 \partial_x^2 C_0 -\left(\partial_x C_0\right)^2\right] \ . 
\end{equation}
The solutions of this differential equation satisfy either
\begin{equation}
 0 = 4 C_0^3 - \left(\partial_x C_0\right)^2
\label{1stFactor}
\end{equation}
or 
\begin{equation}
 0 = C_0 \partial_x^2 C_0 -\left(\partial_x C_0\right)^2 \ .
\label{2ndFactor}
\end{equation}
The first possibility \eqref{1stFactor} implies 
$\partial_x \r_0 = 0 $ through eq.\eqref{delxr0},
giving the trivial solution.
We should therefore focus on the second possibility \eqref{2ndFactor}.

The solution of eq.\eqref{2ndFactor} is given by 
\begin{equation}
 C_0 = \hat{c}_0(u) e^{\k_0(u) x} \ , 
\end{equation}
where $\hat{c}_0(u)$ and $\k_0(u)$ are integration constants 
associated with the integrals with respect to $x$. 
As we are assuming that the $u$-derivatives are of $\mathcal O(\alpha)$, 
it leads to breakdown of the assumption 
if their $u$-derivatives are larger than $\mathcal O(\alpha)$.
However, it is natural to assume that $\hat{c}_0(u)$ and $\k_0(u)$ 
are determined by $a(u)$ so that their derivatives are 
also of $\mathcal O(\alpha)$, just like $\dot{a}(u)$. 
In comparison with the static solution in Sec.\ref{static-solution},
the function $\hat{c}_0(u)$ here corresponds to $c_0/\alpha$,
and $\k_0(u)$ here to $2\sqrt{k c_0/(2a)} \simeq 2\sqrt{c_0/\alpha}$ there.
According to eq.\eqref{c0-estimate-1},
$\hat{c}_0 \simeq 1/(4a^2)$ and $\lam_0 \simeq 1/a$.

Plugging it back into eq.\eqref{delxr0}, 
we solve $\r_0$ as
\begin{equation}
 \r_0 = R_0(u) + \frac{4 \hat{c}_0(u) e^{\k_0(u) x} 
 - \k_0^3(u) x}{4 a(u) \k_0^2(u)} \ , 
 \label{r0eq}
\end{equation}
where $R_0(u)$ is an integration constant. 

The solution \eqref{r0eq} of $\r_0$ can be expanded around $x=0$ as
\begin{align}
 \r_0 &=
 R_0 + \frac{\hat{c}_0}{a \k_0^2} 
 + \frac{4 \hat{c}_0 - \k_0^2}{4 a \k_0} x + \frac{\hat{c}_0}{2 a} x^2 
 + \mathcal O(x^3)
 \nn \\
 &= 
 R'_0
 + \frac{\hat{c}_0}{2 a} \left(x + \frac{4 \hat{c}_0 
 - \k_0^2}{4 \hat{c}_0 \k_0}\right)^2
 + \mathcal O(x^3) \ ,
 \label{xi0-0}
\end{align}
where
\begin{equation}
 R'_0 \equiv
 R_0 + \frac{16 \hat{c}_0^2 + 8 \hat{c}_0 \k_0^2 - \k_0^4}{32 \hat{c}_0 a \k_0^2} \ .
\end{equation}

The $x$-independent part $R'_0$ of $\r_0(u, x)$
can be absorbed by a redefinition of $a$
(see eq.\eqref{rux}).
We shall adopt the convention that $a$ is the minimal value of $\r_0$.
This means that
\be
R_0(u) = 
- \frac{16 \hat{c}_0^2 + 8 \hat{c}_0 \k_0^2 - \k_0^4}{32 \hat{c}_0 a \k_0^2}
\label{cond-R0}
\ee
such that $R'_0 = 0$.

Furthermore,
under the coordinate transformation 
\be
x \rightarrow x + f(u),
\label{x-shift}
\ee
with
\be
f(u) \equiv - \frac{4 \hat{c}_0 - \k_0^2}{4 \hat{c}_0 \k_0}
= - \frac{1}{\k_0} + \frac{\k_0}{4\hat{c}_0} \ ,
\ee
eq.\eqref{xi0-0} becomes
\be
 \r_0(u, x) =
 \frac{\hat{c}_0(u)}{2 a(u)} x^2
 + \mathcal O(x^3) \ .
 \label{xi0-1}
\ee
Therefore,
by a redefinition of $x$,
the local minimum of $\r_0$ appears at $x = 0$ for all $u$.
This is equivalent to say that we can choose
the coordinate system such that
the coefficient functions $\hat{c}_0(u)$ and $\k_0(u)$ satisfy the condition
\begin{equation}
 \hat{c}_0(u) = \frac{1}{4} \k_0^2(u) \ .
 \label{cond-c0}
\end{equation}
While the static solution can be viewed as 
a special case of the dynamical solution for $\beta = 0$,
eq.\eqref{cond-c0} is automatically satisfied by the static solution.

Strictly speaking,
a coordinate transformation \eqref{x-shift}
implies a change in the meaning of 
$\left.\partial_u\right|_x$,
and hence it can in principle lead to
the breakdown of the assumption \eqref{delu=Oalpha},
if $\dot{f}(u)$ is larger than $\mathcal O(\alpha)$.
However, it is expected that $\dot{f}(u)$
is of $\mathcal O(\alpha)$ as for the derivatives 
of $\hat{c}_0(u)$ and $\k_0(u)$. 
In addition,
even the definition of $x$ by \eqref{x-def} can be preserved
by a simultaneous redefinition of the coordinate $u$ as
\be
u \rightarrow u - f(u),
\label{u-shift}
\ee
since $f(u)$ is a function of $u$ only.
The semi-classical Einstein equations in terms of
the coordinates $(u, x)$ thus remain exactly the same
under the simultaneous coordinate transformation
of eqs.\eqref{x-shift} and \eqref{u-shift}.

Finally, 
the difference between \eqref{equu} and \eqref{eqvv} implies that
\begin{equation}
 \dot a(u) = - \frac{\alpha \beta_0(u)}{a(u) \k_0(u)} \ . 
\label{beta-da}
\end{equation}
If $x$ is chosen such that $x=0$ is the local minimum of $r$, 
eq.\eqref{cond-c0} can be used to rewrite it as
\begin{equation}
 \dot a(u) = - \frac{\alpha \beta_0(u)}{2 \hat{c}_0^{1/2}(u) a(u)} \ . 
 \label{dota-2} 
\end{equation}
This equation is in agreement with
eq.\eqref{dota-1} in the previous section.

\subsection{Next-to-Leading Order}

Now, we consider the next-to-leading order terms,
which is of $\mathcal O(\alpha^2)$ in the semi-classical Einstein equation. 
Carrying out calculations in a similar fashion
as we did at the leading order in the previous subsection, 
we first solve for $C_1(u,x)$ and $\r_1(u,x)$ as 
\begin{align}
 C_1 &= 
 e^{\k_0(u) x}\left(\hat{c}_1(u) + \hat{c}_0(u) \k_1(u) x 
 - \frac{2 \hat{c}_0^2(u) e^{\k_0(u)x}}{\k_0^2(u) a^2(u)} 
 + \alpha^{-1} \hat{c}_0(u) \dot \k_0(u) x^2 \right) \ , 
\label{sol-c1}
\\
 \r_1 &= 
 R_1(u) + \frac{\hat{c}_1(u) e^{\k_0(u) x}}{\k_0^2(u) a(u)} 
 - \k_1(u) \left[ \frac{x}{4a(u)} 
 + \frac{\hat{c}_0(u) e^{\k_0(u) x}\left(2 - \k_0(u) x\right)}{\k_0^3(u) a(u)}\right]
\notag\\&\quad 
 - \frac{1}{a^3(u)} 
 \left[
  \frac{\hat{c}_0^2(u)}{\k_0^4(u)} e^{2 \k_0(u) x} 
  - \frac{\hat{c}_0(u) e^{\k_0(u) x}\left(4 + \k_0(u) x\right)}{4 \k_0^2(u)}
  + \frac{\k_0(u) x}{16} - \frac{\k_0^2(u) x^2}{32} 
 \right]
\notag\\&\quad 
 + \frac{2 \dot{\hat{c}}_0(u) e^{\k_0(u) x}}{\alpha \k_0^3(u) a(u)} 
 + \frac{\dot \k_0(u) x \left(\k_0(u) x - 2\right)}{4 \alpha \k_0^3(u) a(u)} 
 \left(4 \hat{c}_0(u) e^{\k_0(u) x} - \k_0^2(u)\right) \ .
\label{sol-r1}
\end{align}
Here $R_1(u)$, $\hat{c}_1(u)$ and $\k_1(u)$ are integration constants.

Again, 
we absorb $R_1(u)$ by a definition of $a(u)$
and impose the condition that 
$r$ has the local minimum at $x=0$
by a simultaneous shift of $x$ and $u$ by a function of $u$.
While this implies eqs.\eqref{cond-R0} and \eqref{cond-c0} at the leading order,
at the next-to-leading order the analogous conditions are 
\begin{align}
 \hat{c}_1(u) &= \frac{1}{2} \k_0(u) \k_1(u) 
 - \frac{\k_0^2(u)}{8 a^2(u)} - \alpha^{-1} \dot \k_0(u) \ , 
\\
 R_1(u) &= - \frac{1}{8 a^3(u)} \ . 
\label{cond-r1}
\end{align}
Furthermore,
since the integration constant $\k_1(u)$ can be absorbed 
by redefining $\k_0(u)$, we can simply take $\k_1(u) = 0$.%
\footnote{
When $\hat{c}_1(u)$ is also absorbed by $\hat{c}_0(u)$, 
the condition \eqref{cond-c0} is modified by $\mathcal O(\alpha)$ terms as
\begin{equation}
 \hat{c}_0(u) = \frac{1}{4} \k_0^2(u) 
 - \frac{\alpha \k_0^2(u)}{8 a^2(u)} - \dot \k_0(u) + \mathcal O(\alpha^2) \ . 
\end{equation}
}

The relation \eqref{beta-da} between the Hawking radiation $\beta(u)$ and
the time dependence of the quantum Schwarzschild radius $a(u)$
can now be improved to including the next-to-leading order terms.
It can be written as 
\begin{equation}
 \beta(u) 
 = 
 - \alpha^{-1} \k_0(u) a(u) \dot a(u) + \frac{\k_0(u) \dot a(u)}{2 a(u)} 
 - \frac{2 a(u)}{\alpha \k_0(u)} \left(\k_0(u) \ddot a(u) - 2 \dot \k_0(u) \dot a(u)\right) 
 + \mathcal O(\alpha^2) \ . 
 \label{eq-dota-2}
\end{equation}
Recall that $u$-derivatives are of $\mathcal O(\alpha)$,
while $a(u)$ and $\k_0(u)$ are of $\mathcal O(\alpha^0)$.
Hence the first term on the right hand side is of $\mathcal O(\alpha^0)$
and the remaining terms are of $\mathcal O(\alpha)$.

In principle, $\beta(u)$ is given by a fixed formula for the Hawking radiation 
that depends on how the initial quantum state evolves in a time-dependent background,
following the choice of a quantum field theoretic model of the vacuum.
Eq.\eqref{eq-dota-2} is the evolution equations for $a(u)$, 
and determines the $u$-dependence of $a(u)$ for a given Hawking radiation.

\subsection{Apparent Horizon}

Let us now study the apparent horizon of the geometry. 
The apparent horizon is the boundary of the trapped region in which 
both inward- and outward-pointing null vectors are converging.
That is,
\begin{align}
 \left(\frac{\partial r}{\partial v}\right)_u &< 0 
& &\text{and} &
 \left(\frac{\partial r}{\partial u}\right)_v &< 0 \ ,
\label{cond-tr}
\end{align}
where the subscripts $u$ and $v$ outside the parentheses
refer to the variables fixed for the partial derivatives.
Using eqs.\eqref{rux}, \eqref{r0eq}, \eqref{cond-R0} and \eqref{cond-c0}, 
we can expand $r$ as 
\begin{equation}
 r(u,x) = 
 a(u) + \alpha\frac{e^{\k_0(u) x} - (1 + \k_0(u) x)}{4 a(u)} 
 + \mathcal O(\alpha^2) \ .
\end{equation}
Then the conditions \eqref{cond-tr} become 
\begin{equation}
 \left(\frac{\partial r}{\partial v}\right)_u = 
 \frac{\alpha \k_0(u) \left(e^{\k_0(u) x} - 1\right)}{8 a(u)} 
 + \mathcal O(\alpha^2) 
 \ < \ 0 
\end{equation}
and 
\begin{equation}
 \left(\frac{\partial r}{\partial u}\right)_v
 = \dot a(u) - \frac{\alpha \k_0(u) \left(e^{\k_0(u) x} - 1\right)}{8 a(u)} + \mathcal O(\alpha^2) \ < \ 0 \ . 
\end{equation}
The boundary of this region is given by 
\begin{align}
 \left(\frac{\partial r}{\partial v}\right)_u &= 0 
& &\text{or} &
 \left(\frac{\partial r}{\partial u}\right)_v &= 0 \ , 
\end{align}
which are solved to give the $x$-coordinates of
the outer and inner boundaries as
\begin{equation}
 x_O = 0 + \mathcal O(\alpha) \ ,
\label{outAH}
\end{equation}
and 
\begin{equation}
 x_I = \frac{1}{\k_0(u)} \log\left(1 + \frac{8 a(u) \dot a(u)}{\alpha \k_0(u)}\right) + \mathcal O(\alpha) 
 = \frac{1}{\k_0(u)} \log\left(1 - \frac{8 \beta(u)}{\k_0^2(u)}\right) + \mathcal O(\alpha) \ , 
\label{inAH}
\end{equation}
respectively. 
Since $\dot a$ is negative, 
or equivalently, $\beta$ is positive, 
\eqref{inAH} implies that $x_I < 0$. 

One can check that the conditions \eqref{cond-tr} are satisfied 
in the region where $x_I < x < x_O$.
Hence \eqref{outAH} is the outer apparent horizon and 
\eqref{inAH} is the inner apparent horizon. 
The outer apparent horizon coincides with the local minimum of $r$ on $u=$ const.\ surfaces, 
which can be chosen to be located at $x=0$ to all orders
by a shift of $x$ \eqref{x-shift}.

The inner apparent horizon coincides with the outer apparent horizon
when there is no Hawking radiation ($\beta = 0$).
In this case there is no trapped region 
and therefore strictly speaking no apparent horizon.
This is consistent with the claim \cite{Ho:2017joh} that
there is no apparent horizon for the static solution.

It should be noted that the geometry described above
includes the quantum effect of the vacuum energy-momentum tensor
but does not have the effects of the collapsing matters,
as we are focusing on the space-time region outside the collapsing star.
In general,
in order to consider the apparent horizon for
the physical process of black-hole evaporation, 
the geometry inside the collapsing matter should be taken into account. 

The simplest model of a collapsing star
is a spherically symmetric thin shell.
The curved geometry outside the shell meets the interior flat spacetime on the shell.
From both the viewpoints of the interior and exterior space-time,
the radius $r$ of the shell decreases with time,
due to the junction condition on the shell. 

For a thin shell collapsing at the speed of light,
its trajectory is a constant-$v$ curve.
Since the condition 
\begin{equation}
 \left(\frac{\partial r}{\partial u}\right)_v < 0 
\end{equation}
should be satisfied on the shell collapsing at the speed of light,
the shell must be outside of the inner apparent horizon \eqref{inAH}
in the exterior geometry above, 
or equivalently, 
\begin{equation}
 x_S > x_I \simeq \frac{1}{\k_0(u)} \log\left(1 - \frac{8 \beta(u)}{\k_0^2(u)}\right) \ 
\end{equation}
for the $x$-coordinate of the shell denoted by $x_S$.
If $x_S > x_O$,
i.e. the $x$-coordinate of the shell is larger than
the $x$-coordinate of the outer apparent horizon,
it would mean that there is no apparent horizon.
If, on the other hand,
$x_S < x_O$, 
the outer horizon does exist at $x_O$.
The region with $x \in (x_S, x_O)$ is then part of the trapped region,
while the internal space of the shell ($x < x_S$) is not 
part of the trapped region since it is flat. 
The inner apparent horizon, 
which is the inner boundary of the trapped region, 
must therefore coincide with the shell. 

For a thin shell collapsing at a speed slower than light,
however,
it is possible that 
there is an inner apparent horizon outside the collapsing shell.

\section{Geometry of Evaporating Black Holes}

Through the calculation above,
one can deduce the geometry of the dynamical process of
black-hole evaporation for the model of vacuum energy-momentum tensor
given in eqs.\eqref{Tuu-vac} -- \eqref{Tthth-vac}.
A salient feature of the evaporation process is 
the presence of a wormhole-like structure near the Schwarzschild radius.
This feature is in fact robust whenever 
there is ingoing negative energy flux.
We claim that,
in any model of vacuum energy-momentum tensor,
whenever there is a negative energy flux so that
\begin{equation}
T_{vv}^{(hor)} \equiv 
\left.\langle T_{vv} \rangle\right|_{v = v_0(u)} < 0
\end{equation}
at the apparent horizon,
the apparent horizon must also be a local minimum of 
the areal radius.
Here $v_0(u)$ is the $v$-coordinate of the apparent horizon.

This statement can be easily proved as follows.
The semi-classical Einstein equation $G_{vv} = \kappa \langle T_{vv} \rangle$ 
says that
\begin{equation}
\frac{2\del_v C\del_v r}{Cr} - \frac{2\del_v^2 r}{r}
= \kappa T_{vv}.
\label{EE-vv-hor}
\end{equation}
When the collapsing matter is under the apparent horizon,
$T_{vv}$ is given by that of the vacuum $\langle T_{vv} \rangle$.
At the apparent horizon,
$\del_v r = 0$,
it reduces to 
\footnote{
An exception is when $C = 0$ at the apparent horizon.
This happens for the Schwarzschild solution
in which the apparent horizon is also an event horizon.
In a more realistic solution,
we have $C \neq 0$
at the apparent horizon.
}
\begin{equation}
\del_v^2 r = - 2 \kappa a(u) T_{vv}^{(hor)} > 0,
\end{equation}
where $a(u)$ is the value of $r$ at the apparent horizon.
This implies that the areal radius $r$ can be expanded as
\begin{align}
r(u, v) &=
a(u) + \kappa a(u) \left|T_{vv}^{(hor)}\right| (v - v_0(u))^2 + \cdots
\nn \\
&= a(u) + 4 \kappa a(u) \left|T_{vv}^{(hor)}\right|
\left(x - x_0(u) \right)^2 + \cdots,
\end{align}
where
$x_0(u) \equiv \frac{2v_0(u) - u}{2}$,
around the apparent horizon at $v = v_0(u)$.
It is thus clear that the apparent horizon is a local minimum of $r$.

This simple calculation also shows when and where
the apparent horizon first appears.
Before the apparent horizon appears,
there is no trapped region,
and the energy-momentum tensor includes not only 
the energy-momentum tensor $\langle T_{vv} \rangle$ of the vacuum
but also that of the collapsing matter.
If there is a trapped region at later times,
the boundary of the trapped region is the apparent horizon.
Let us now look for the point on the apparent horizon 
that has the minimal value of $u$,
say, $u = u_0$.
This is the point where the apparent horizon first appears.
Assuming that the apparent horizon is a smooth subspace,
we have 
\begin{equation}
\del_v r = \del_v^2 r = 0
\end{equation}
at this point $(u = u_0, v = v_0(u_0))$.
Hence eq.\eqref{EE-vv-hor} implies that
\begin{equation}
\left. T_{vv}\right|_{v = v_0} = 0.
\end{equation}
This happens when the ingoing negative energy flux 
$\langle T_{vv} \rangle$ of the vacuum 
happens to cancel the positive energy of the collapsing matter.
Typically,
the collapsing matter has a density distribution that decays to zero
at the surface of the collapsing star,
and the exact cancellation happens very close to the surface of the star
(but inside the star)
where the classical energy flux is as weak as the quantum vacuum energy flux in magnitude.

As the matter continues to collapse
under the neck of the wormhole-like structure,
the total energy of the star decreases with time
due to Hawking radiation.
The Schwarzschild radius $a_0$ (and $a$) 
eventually approaches to the Planck scale
where the low-energy effective description is no longer valid.

\begin{figure}
\vskip1cm
\begin{center}
\includegraphics[scale=0.45,bb=0 90 330 180]{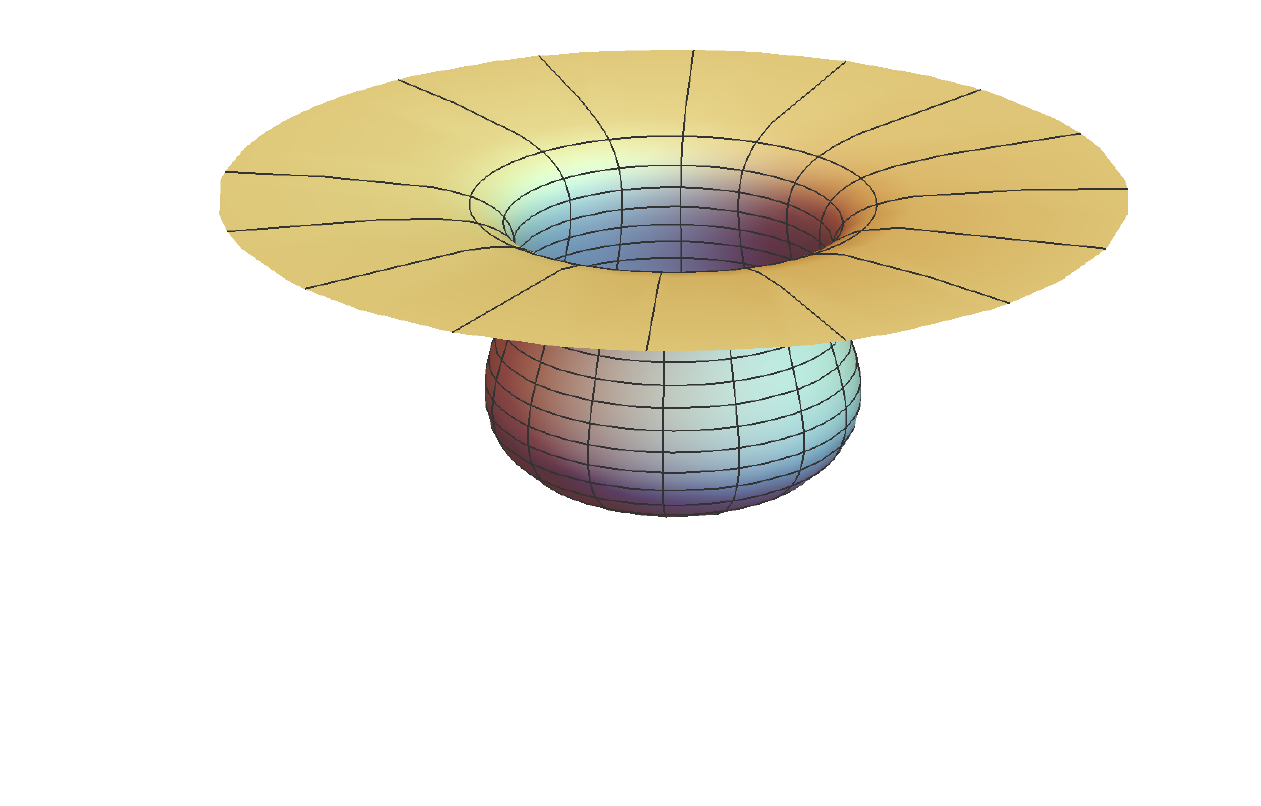}
\includegraphics[scale=0.45,bb=0 90 330 180]{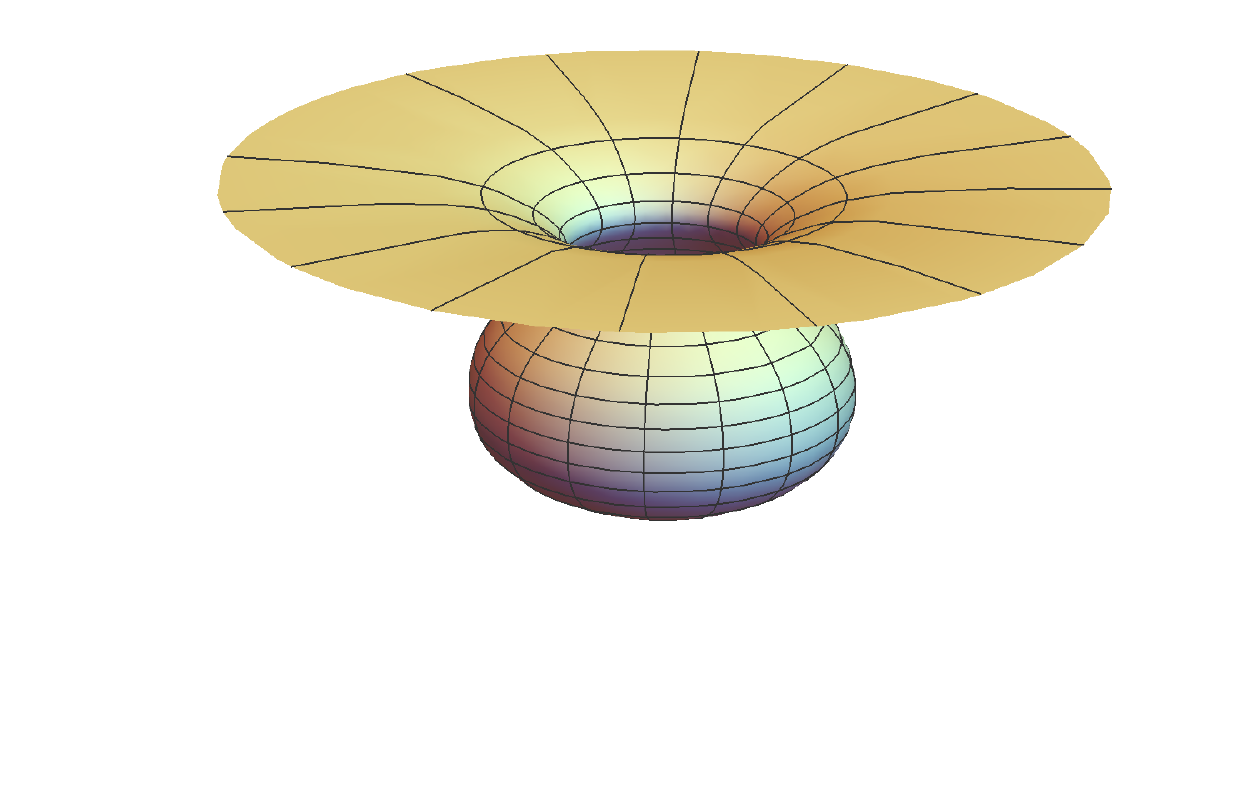}
\includegraphics[scale=0.45,bb=0 90 330 180]{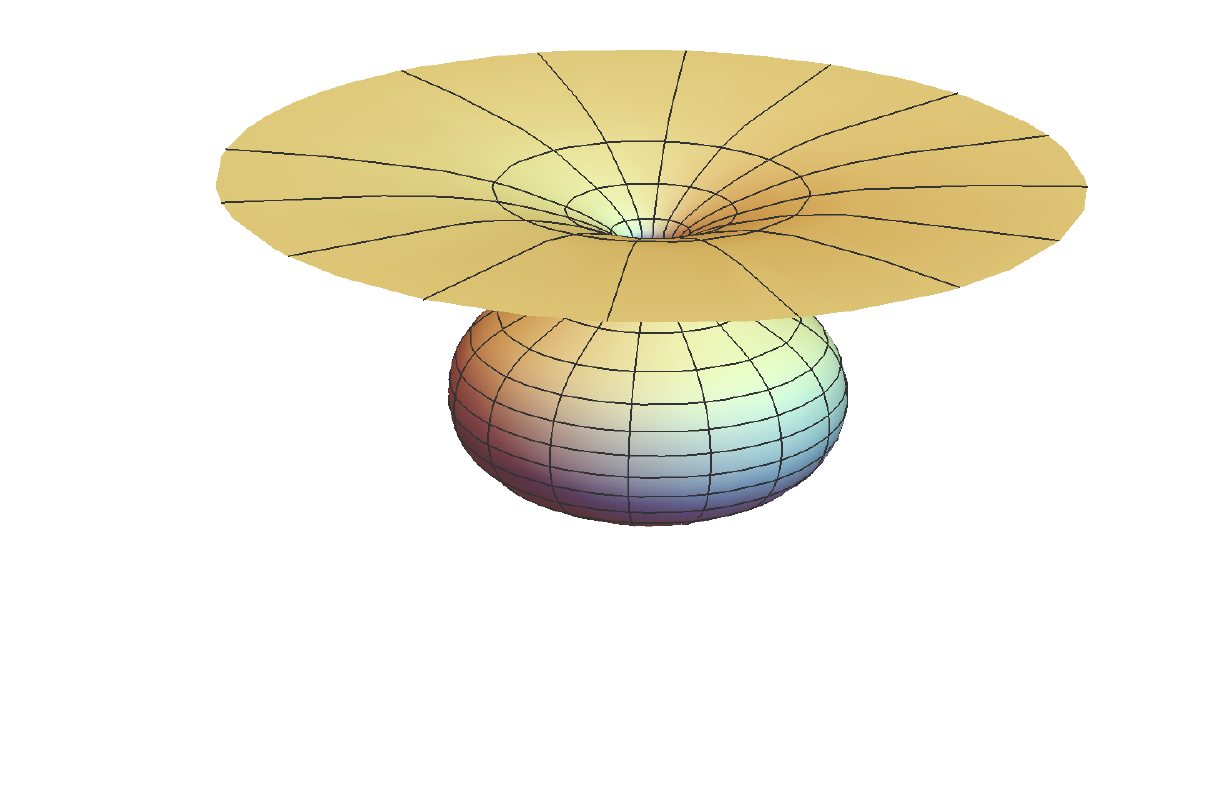}
\caption{\small 
The schematic diagrams for an evaporating black hole
that ends up with a Planckian neck and a large internal space,
from left to right in time sequence.
Note that the notion of distance is distorted in the diagrams
because this geometry does not admit an embedding in 3D flat space.
}
\label{figure:WH1}
\end{center}
\end{figure}

When the neck shrinks to the Planck scale,
the size of the internal space under the neck
can be either of macroscopic scale or of Planck scale.
In Ref.\cite{Parentani:1994ij},
it was found that,
for a theory of massless scalar field,
the internal space is still large when
the neck shrinks to a Planckian size.
The evolution history of this scenario is shown 
schematically in Fig.\ref{figure:WH1}.
The final state of a large internal space under 
a small neck resembles the geometry of 
the so-called Wheeler's bag of gold \cite{Wheeler-Bag-Of-Gold}.

\begin{figure}
\vskip1cm
\begin{center}
\includegraphics[scale=0.45,bb=0 90 330 180]{BH-1.pdf}
\includegraphics[scale=0.45,bb=0 65 330 155]{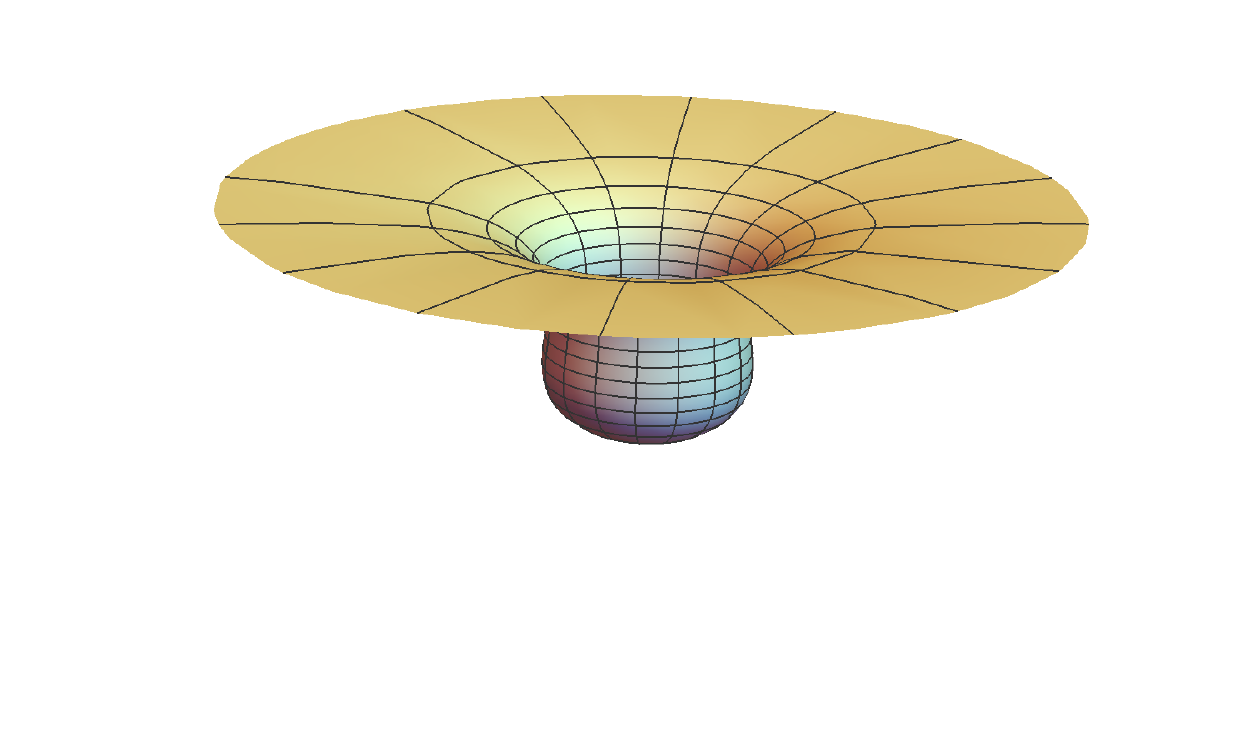}
\includegraphics[scale=0.42,bb=0 -20 330 60]{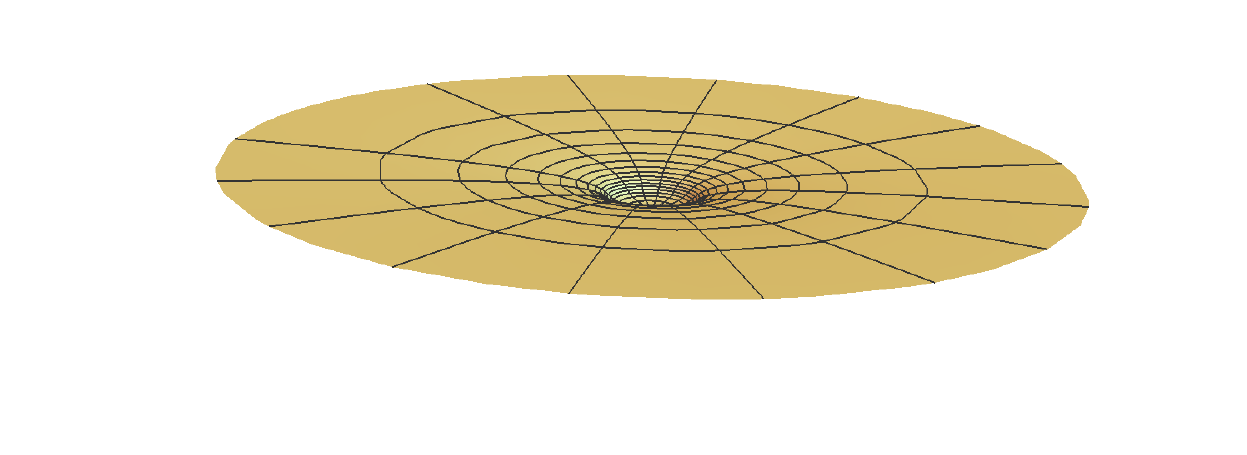}
\caption{\small 
The schematic diagrams for an evaporating black hole
that ends up with a Planck-scale internal space,
from left to right in time sequence.
Note that the notion of distance is distorted in the diagrams.
}
\label{figure:WH2}
\end{center}
\end{figure}

The other scenario in which the neck and the internal space
shrink to zero can be realized as follows.
Take the static solution in Refs.\cite{Ho:2017joh,Ho:2017vgi}
and tune the mass parameter adiabatically to zero.
The size of the internal space is thus correlated with
the size of the neck,
and the scale of the black hole shrinks to a Planckian size as a whole.
This is shown schematically in Fig.\ref{figure:WH2}.

\section{Discussions}
\label{Conclusion}

In this work,
we study the geometry of the collapse of compact stars 
with the back reaction of the vacuum energy-momentum tensor included.
The vacuum energy-momentum tensor is assumed to be 
qualitatively approximated by a simple 2D quantum field theory
proposed in Ref.\cite{Davies:1976ei}.
A crucial feature of this model is the presence of the negative vacuum energy.
The conclusion of this work therefore cannot apply to models 
in which the vacuum energy is non-negative.
For instance, 
the black-hole models considered in
Refs.\cite{Kawai:2013mda}-
-\cite{Ho:2015vga}, \cite{Ho:2016acf,Kawai:2017txu}
demonstrate a different nature.

The static geometry for this model as non-perturbative solutions
to the semi-classical Einstein equations
was previously obtained in Refs.\cite{Ho:2017joh,Ho:2017vgi}.
At large distances outside the Schwarzschild radius,
the Schwarzschild solution is a very good approximation.
But when one gets very close to the Schwarzschild radius
(when the separation is of the Planck scale),
non-perturbative effects are found to result in crucial deviations 
from the Schwarzschild solution.
The exact static solution has no horizon and Buchdahl's theorem is circumvented,
with the horizon replaced by a wormhole-like geometry,
i.e.\ a local minimum of the areal radius $r$.
In this paper,
we turn on Hawking radiation and 
study the time-dependent perturbation theory
to describe the gravitational collapse of
a black-hole-like object in this model.

In the dynamical case,
the Schwarschild solution with a time-dependent parameter $a_0(u)$
is still expected to be a good approximation at large distances.
We focus our attention on the geometry around the neck 
of the wormhole-like structure,
assuming that the collapsing matter is already under the neck.
We found a time-dependent solution in which
the neck coincides with an apparent horizon $a(u)$,
which is approximately equal to 
the (classical) Schwarzschild radius $a_0(u) = 2M(u)$.

Due to Hawking radiation,
the radius of the neck $a(u)$ decreases with time,
and the wormhole-like structure is found to persist.
There are two possible scenarios depicted 
in Fig.\ref{figure:WH1} and Fig.\ref{figure:WH2},
depending on whether the size of the interior space under the neck 
shrinks together with the neck.

Numerical simulation \cite{Parentani:1994ij} showed that
the scenario depicted in Fig.\ref{figure:WH1} is realized 
for a fast collapse 
in which the energy-momentum tensor is approximated by 
that for 2D massless scalar fields including the collapsing matter. 
On the other hand,
the scenario depicted in Fig.\ref{figure:WH2} 
can be realized by an extremely slow collapse as an adiabatic process 
defined by slowly tuning the mass of the static solution of 
Ref.\cite{Ho:2017joh,Ho:2017vgi}.

When the radius $a(u)$ of the neck approaches to the Planck size,
the low-energy description is no longer valid,
there are two possibilities for the fate of the black hole
depending on the UV physics.
One possibility is that the neck remains finite $a(u) > 0$
so that the interior space under the neck remains
attached to the exterior space.
All the information inside the bag remains accessible to exterior observers.
There is no event horizon in this case.
The other (more dramatic) possibility is that,
through a quantum transition,
the interior space is detached from the exterior spacetime,
and the singularity of detachment, 
although it is expected to modified by the UV physics of gravity, 
marks the boundary of an event horizon. 

Let us now comment on the information loss paradox
\cite{Hawking:1976ra,Mathur:2009hf,Marolf:2017jkr}.
The vacuum energy-momentum tensor adopted in this model
has negative energy flowing into the Schwarzschild radius from the outside,
and that is why the total energy under the neck appears to decrease over time.
This is the same as the conventional model of black holes,
but it does not immediately imply that it is impossible for
the Hawking radiation to carry information of the collapsing matter.
The outgoing energy flux $T_{uu}$ \eqref{Tuu-beta} consists of two parts.
The first part is an outgoing negative energy flow. 
This is a part of the negative vacuum energy and 
is balanced with an ingoing negative energy flow $T_{vv}$,
in the static case. 
The second part involving $\beta(u)$ is the Hawking radiation.
For 2D scalar fields, 
the energy-momentum tensor is calculated 
by using the conservation equation and Weyl anomaly. 
The first part corresponds to a special solution of the conservation equation. 
It is directly related to the Weyl anomaly and can be chosen such that 
it is independent from the physical states of matter fields.
Therefore, the first part has no contribution to entropy. 
The second part comes from the integration constants
for the conservation equation, 
and should be determined by the initial condition. 
It depends on physical states of matters and can contribute to entropy. 
Therefore, Hawking radiation can
carry information from the collapsing star.
In fact, $\beta$ is non-zero at the collapsing star
even when $T_{uu}$ is zero there. 

However, the Hawking radiation cannot
carry all information about the collapsing matter,
unless it has direct interaction with the collapsing matter.
It is determined by the initial state of the quantum field of Hawking radiation
as well as the geometry of space-time in its evolution history.
While the space-time geometry is sourced by the collapsing matter,
the former is only sensitive to the energy-momentum tensor,
but not to other details of the latter.
In fact,
even when there is direct interaction between 
the quantum field of Hawking radiation
and the collapsing matter,
there cannot be complete information transfer
in a low energy effective theory.
(This is easy to see when there is a global symmetry.)
For Hawking radiation to carry all information 
about the collapsing matter,
we need something to occur at the Planck-scale
to invalidate the low energy effective theory
(i.e., a violation of a certain ``niceness condition''
\cite{Mathur:2009hf}),
so that a UV theory such as string theory is a necessity.
When the collapse can be described as a small perturbation
of the static configuration,
there is indeed a Planck-scale 
energy density and pressure under the surface of the collapsing matter,
which is only a Planck-scale distance away from the neck
\cite{Ho:2017joh,Ho:2017vgi}.
The information loss paradox is thus no longer
a low-energy effective theoretical problem.

\section*{Acknowledgement}

The author would like to thank 
Emil Akhmedov,
Jiunn-Wei Chen,
Chong-Sun Chu,
Hsien-chung Kao,
Hikaru Kawai,
Samir Mathur,
Andrei Mironov,
Alexei Morozov,
Yu-Ping Wang,
Shu-Jung Yang and Yuki Yokokura
for discussions.
The work is supported in part by
the Ministry of Science and Technology, R.O.C.
(project no. 104-2112-M-002 -003 -MY3)
and by National Taiwan University
(project no. 105R8700-2).


\end{document}